# Characteristics study of projectile's lightest fragment for $^{84}Kr_{36}$ - emulsion interaction at around 1 A GeV

N. Marimuthu[1, 2], V. Singh[1]*, S. S. R. Inbanathan[2]

*1 High Energy Physics laboratory, Physics Department, Banaras Hindu University, Varanasi-221005, INDIA*
*2 The American College, Madurai-625002, Tamilnadu, INDIA*

*E-mail: *venkaz@yahoo.com*

**Abstract** : The present article significantly investigated projectile's lightest fragments (proton) multiplicity distribution and probability distribution with $^{84}Kr_{36}$ emulsion collision at around 1 A GeV. The multiplicity and normalized multiplicity of projectile's lightest fragments (proton) is correlated with the compound particles, shower particles, black particles, grey particles, helium fragments particles and heavily ionizing charged particles. It is found that projectile's lightest fragments (proton) are strongly correlated with compound particles and shower particles rather than other particles and the average multiplicity of projectile's lightest fragments (proton) increases with increasing compound, shower and heavy ionizing particles. Normalized projectile's lightest fragments (proton) are strongly correlated with compound particles, shower particles and heavy ionizing charge particles. The multiplicity distribution of the projectile's lightest fragments (proton) emitted in the $^{84}Kr_{36}$ + emulsion interaction at around 1 A GeV with different target has well explained by KNO scaling. The mean multiplicity of projectile's lightest fragments (proton) depends on the mass number of the projectile and does not significantly dependent of the projectile energy. The mean multiplicity projectile's lightest fragment (proton) increases with increasing the target mass number.

## 1. INTRODUCTION

Nowadays experimental high energy physicists are trying to understand the Multifragmentation process and interaction with target to provide some aspects of nuclear structure [1-2]. Nuclear emulsion detector (NED) technique is a very important tool for intermediate and high energy physics. This is understood through a long list of fundamental discoveries. For the last several years, nuclear emulsion techniques used to investigate nucleus - nucleus and nucleus-hadrons collisions. Because of the nuclear emulsion detector provides $4\pi$ geometry, submicron spatial resolution and excellent detection of relativistic particles. In the past few decades' large number of papers had investigated the multiplicity distribution and multiplicity correlation with evaporated and recoil nucleons of target, produced mesons, helium and heavily ionizing charged particles [3-9]. It is believed that, it may help us to understand the interaction mechanism and rare particle production. The fragmentation has been one of the most important aspects of nucleus-nucleus collision and hadrons-nucleus collision. According to the participant-spectator model [10-13], heavy ion interaction regions are called as participant region and rest regions are spectators. During collision, in the participant region we may expect the local density and temperature to increase and after that the participant region expands and cools down. The single charged relativistic particles and rare isotopes are formed from participant region. These are mostly mixture of pions, k mesons and less proportional to the fast protons. The target fragment is formed from the target spectator. The target fragment is the combining of recoiled proton and evaporated fragments. According to the emulsion detector terminology, the target recoiled particles are called as grey particles and evaporated target fragments are black particles. The projectile fragments formed from the highly excited projectile spectator residues through the evaporation. The projectile fragment is speculated, because it is coming from highly excited projectile spectator nuclei and the projectile's lightest fragment (proton) is one of the parts of projectile fragment. It may carry the information about the equation of state, dynamical mechanism and liquid phase transition of low density matter [2, 14]. So it is very important to investigate projectile's lightest fragments (proton). In this paper, we have focused on complete investigation of the projectile's lightest fragments (proton) multiplicity and its correlation with other secondary particles of $^{84}Kr_{36}$ with NIKI-BR emulsion plates at kinetic energy around 1 GeV per nucleon.

## 2. EXPERIMENTAL DETAILS

We have analyzed the data which were collected from the nuclear emulsion detector experiment. We used highly sensitive stack of NIKI-BR2 emulsions plates having unprocessed thickness of 600 μm. These emulsion plates were exposed horizontally with the beam of $^{84}Kr_{36}$ having kinetic energy around 1 GeV per nucleon at the edge of the plate. This experiment has been performed in Gesellschaft fur Schwerionenforschung (GSI) Darmstadt, Germany. The events / interactions observed through the line scanning technique with oil immersion objective lens of 100X along with the 15X eyepieces. There are two types scanning methods used to study the emulsion plates; one was line scanning and the other one volume scanning. In the line scanning method the beam tracks are picked up at 5mm distance from the edge of emulsion plate and it is carefully followed, since they are interacting with the nuclear emulsion nuclei or escaped from any sides of the emulsion plate. To make sure that we are following primary track first we followed the track in the backward direction till the edge of the plate. In the volume scanning methods, the information is collected through strip by strip scanning [6-8]. The first interaction / event observed are the primary interaction and emitted particles are the secondary one. According to the emulsion terminology, the emitted secondary charged particles are classified through their range, velocity and ionization and some of the following categories.

**Black particles**- The black particles are evaporated target fragments. Most of them are produced from the residue of target nuclei and their particles range < 3mm. The relativistic velocity of the particles are 0.3c and their kinetic energy < 20MeV. The multiplicity of black particles is denoted as $N_b$.

**Grey particles-** These particles are mostly recoil proton of target and their range should be greater than 3mm. These particle have relativistic velocity $0.3c < \beta < 0.7c$ and their kinetic energy $30 < E < 400$ MeV. The multiplicity of these particles is denoted as $N_g$.

**Shower particles-** These particles are singly - charged freshly created relativistic particle. These particles come from the participant region and their velocity is greater than $0.7c$. Mostly shower particles are contaminated with pions, kaons and less number of fast protons. The multiplicity of shower particles is denoted as $N_s$.

A set of these particles form a new parameter called compound particles defined as a sum of recoil proton and shower particles and their multiplicity denoted as $N_c = (N_g + N_s)$.

**Heavily Ionizing charged particles-** These charged particles are the part of target nucleus or group of the target fragments. It is equal to sum of the black and gray particles $N_h = (N_b + N_g)$. The multiplicity of heavily ionizing charged particles denoted as $N_h$.

**Projectile fragments-** The projectile fragments are spectator part of the projectile nucleus or projectile spectators. In the present work, the projectile fragments have been categorized on the basis of their charge as follows:

**Singly charged projectile fragments or projectile's lightest fragments (proton) -** These are mostly projectile proton ($PF_{Z=1}$) and velocity of this particle is almost equal to the beam velocity. It is denoted as $N_p$.

**Alpha projectile fragments-** These particles are double charged projectile fragments ($PF_{Z=2}$). It is denoted as $N_{Alpha}$.

**Heavy projectile Fragments-** These particles are multiple charged projectile fragments ($PF_{Z<2}$). It is denoted as $N_f$.

## 3. EXPERIMENTAL RESULTS AND DISCUSSION

In the present paper, we have investigated 892 primary inelastic events chosen from the $^{84}Kr_{36}$ - emulsion interaction at around 1 A GeV. Initially we carried out the target identification for $^{84}Kr$-emulsion interactions, since the exact target identification is not possible in case of NED because the nuclear emulsion medium is composed of the H, CNO and Ag / Br compounds and molecules. However we have classified target groups statistically on the basis of target projectiles multiplicity for each event. The $N_h \leq 1$ events taken to be $^{84}Kr$ - H interaction, events with $2 \leq N_h \leq 7$ are $^{84}Kr$ - CNO interactions and events having $N_h \geq 8$ are $^{84}Kr$ – Ag/Br interactions. The number of events corresponding to these interactions is 136, 300 and 456, respectively. Multiplicity distribution of projectile's lightest fragments (proton) produced in $^{84}Kr_{36}$ - emulsion at around 1 A GeV interactions is shown in Figure 1. Here the target Ag / Br and $^{84}Kr$ - emulsion interactions distributions are fitted with the Gaussian function. The distribution maximum values are extended up to 20 for both the distributions. The distributions of projectile's lightest fragments (proton) shapes are almost similar for different targets, whereas the distribution becomes wider with increase of target mass. Table 1 show the Gaussian distribution parameters value for different targets.

Table 2 shows the mean multiplicity of the projectile's lightest fragments (proton) for different target groups of $^{84}Kr$ - emulsion interactions at around 1 A GeV and for comparing other projectiles with similar and different energies results are also listed. From table 2 one may see that the mean multiplicity of projectile's lightest fragments (proton) for different target groups increases with increasing the average target mass and it decreases by increasing the projectile kinetic energy.

Figure 2 depicted the probability distribution of projectile light fragment (proton) fitted with the Gaussian distribution function. The maximum value of probability is found to be 21.74±2.2 with the peaks and the width of the Gaussian function is -10.07 and 16.28, respectively. The probability distribution of projectile light fragment (proton) extends up to 20 in an event.

Data presented in the table 3 shows the average multiplicities of projectile's light fragment (proton) for different projectiles of similar and different kinetic energy. These data reflect that the mean multiplicity of the projectile's light fragment (proton) completely depend on mass number of the projectile, but it does not depend on energy of the projectile because the mean multiplicity of projectile's light fragment (proton) decreases with increasing projectile energy. We studied the dependence of average multiplicity of projectile's light fragment (proton) $<N_p>$ with the average mass number of the target groups ($A_T$) as shown in the Figure 3. It may be seen from Figure 3 that the average multiplicity of projectile's light fragments (proton) increases with increase of target mass. These values are fitted with the linear best fit relation and the value of the fit parameter are given below

$$<N_p> = (3.1514 \pm 0.1758) \, A_T^{(0.3337 \pm 0.1239)}. \tag{1}$$

**3.1 KNO SCALING FOR PROJECTILE'S LIGHT FRAGMENT (PROTON) $<N_p>$:**

The Koba – Nielson - Olesen (KNO) has been one of the dominant frame works to study the behavior of multiplicity distribution of secondary particles in nucleus - nucleus and hadrons-nucleus collisions [24-25]. The KNO scaling hypothesis derived from the Feynman scaling of inclusive particle production cross section. According to the KNO scaling

$$\psi(Z) = 4Z \exp(-2Z), \tag{2}$$

Where

$$\psi(Z) = <N_p> P(N_p) = <N_p> \sigma_{np}/\sigma_{inels.} \tag{3}$$

Here $\psi(Z)$ is the energy independent function $Z = N_p / <N_p>$. Where $N_p$ refers to the event by event produced of projectile's light fragments (proton) and $<N_p>$ represent the average or mean multiplicity of projectile's light fragments (proton) for the whole data sample. $\sigma_{np}$ is referred to the partial cross section of the produced charged particle of projectile's light fragments (proton) multiplicity ($N_p$) for the specific channel and $\sigma_{inels}$ is the total cross section. We know that $\psi$ is energy independent, so it is approximately constant for all beam of same energy. In Figure 4 we have plotted the multiplicity distribution $<N_p>\sigma_{np} / \sigma_{inels}$ versus the $N_p / <N_p>$ for different targets such as $^{84}$Kr - H, $^{84}$Kr - CNO, $^{84}$Kr - AgBr and $^{84}$Kr - emulsion interactions at around 1 A GeV. And here the experimental data point $N_p = 0$ has been excluded. The experimental data are fitted with the universal function $\psi(Z) = AZ \exp^{(-BZ)}$. (4)

From Figure 4 one can see that most of the experimental data are laid on the universal curve and complete range of the spectrum is fitted well within the experimental error. The values of best fitting parameters A and B are obtained and are tabulated in Table 4. From the Table 4, the fitting values of A and B are agreeing well with the theoretical one within the statistical error i.e. A = 4 and b = 2. The KNO fitting parameters of $^{84}$Kr$_{36}$ - emulsion interactions at around 1 A GeV for different target values are compared with the $^{84}Kr_{36}$ - emulsion interactions at 1.7 A GeV as listed in the Table 4 too.

### 3.2 MULTIPLICITY CORRELATION OF $<N_h>$, $<N_b>$, $<N_g>$, $<N_s>$, $<N_{Alpha}>$, $<N_c>$ WITH $N_p$:

The projectile's lightest fragments (proton) are coming from highly excited projectile residue and it might carry information about the collision dynamics. Up to now very less number of work were carried out with projectile's lightest fragments (proton) [15-23]. Figure 5 depicts the multiplicity correlation between secondary particles $<N_j>$ = $<N_h>$, $<N_b>$, $<N_g>$, $<N_s>$, $<N_{Alpha}>$, $<N_c>$ and the number of projectile's lightest fragments (proton) ($N_p$) for $^{84}Kr_{36}$-emulsion interaction at around 1 A GeV. The symbols are experimental data with statistical error-bar and it is fitted with straight line function as mentioned in Eqn. (5) and the fitting parameter values are tabulated in the Table 7. Figure 5 reveals that, the average multiplicity of compound particle ($<N_c>$) and shower particle multiplicity have strong correlation with the number of projectile's lightest fragments (proton) emitted whereas the average multiplicities of black particles, grey particles, alpha particles, and heavy ionizing charged particles have shown weaker correlation with the number of projectile's lightest fragments (proton) i.e. $N_p$ emitted in an event. It may also show that the average multiplicity of compound particle and shower particles are rapidly increasing with increase in the number of projectile's lightest fragments (proton). It shows that the average multiplicity of recoil protons ($N_g$) and evaporated ($N_b$) target fragments are weakly dependent on the number of the projectile's lightest fragments (proton) which are coming from the participant region. The emission rate of the shower particles has strong dependence on the number of projectile's lightest fragments (proton) showing almost similar source of origin i.e. participant region. The maximum number of compound multiplicity and shower particles emitted in an event is 15 and 21, respectively. The other particles like black, grey, alpha and heavy ionized charged particles have not shown significant dependence on the projectile's lightest fragments (proton).

$$<N_j> = b<N_p> + a. \tag{5}$$

## 3.3 MULTIPLICITY CORRELATION BETWEEN OF $<N_p>$ WITH $N_c$, $N_s$, $N_{Alpha}$, $N_b$, $N_g$, AND $N_h$:

The multiplicity correlation of $<N_p>$ with the corresponding $N_c$, $N_s$, $N_{Alpha}$, $N_b$, $N_g$, and $N_h$ for $^{84}Kr_{36}$ + emulsion interactions at around 1 GeV per nucleon are shown in Figures 6 & 7. Here all the distributions are fitted with straight line function and fitting parameters values are given in equation from Eqns. (6) to (9). From the Figure 6, the average number of projectile's lightest fragments (proton) i.e. $<N_p>$ linearly increases with increasing number of shower particles ($N_s$). The $<N_p>$ value rapidly increases with increase in the number of compound particles ($N_c$) up to the 21 $N_c$ per event and after that $N_c$ values have larger error due to less number of events in that $N_c$ region. The $<N_p>$ value initially increases with increasing the number of alpha particles in an event ($N_{Alpha}$). From Figure 7 we may see that the $<N_p>$ value rapidly increases with increasing number of black particles ($N_b$), grey particles ($N_g$) and heavy ionizing charge particles ($N_h$) up to 11 particles per event. It may also be seen that, the $<N_p>$ values are saturated for ($N_h$) values beyond the 13 number of heavily ionizing charged particle. The plot also indicates that heavily ionizing particle and grey particles shows similar strength of the correlation with projectile's lightest fragments (proton). The slopes value for $N_h$ and $N_g$ are (0.13±0.04) and (0.13±0.007), respectively. The maximum number of $N_p$ observed in an event is 36, which is equal to the total charge of the projectile ($^{84}Kr_{36}$). The maximum number of $N_c$ and $N_s$ observed in an event are 31 and 35, respectively. In the case of $N_b$, $N_g$ and $N_h$ the $<N_p>$ values are started from 2.95± 0.24, 2.91±0.10 and 2.52±0.14, respectively.

$<N_p> = (0.17 \pm 0.02) N_s + (1.44 \pm 0.44)$ (6)

$<N_p> = (0.10 \pm 0.02) N_c + (1.93 \pm 0.43)$ (7)

$<N_p> = (0.41 \pm 0.34) N_{Alpha} + (2.12 \pm 1.96)$ (8)

$<N_p> = (0.13 \pm 0.04) N_h + (2.43 \pm 0.50$ (9)

$<N_p> = (0.24 \pm 0.06) N_b + (2.41 \pm 0.43)$ (10)

$<N_p> = (0.13 \pm 0.007) N_g + (3.15 \pm 0.51)$ (11)

## 3.4 NORMALIZED MULTIPLICITY CORRELATION BETWEEN $<N_p>/N_p$ WITH $N_s$, $N_c$, $N_{Alpha}$, $N_h$, $N_b$, AND $N_g$:

Figure 8 (a) and (b) depicted the correlation relation between normalized multiplicity of projectile's lightest fragments (proton) i.e. $(1/N_p)*(d<N_p>/dN_i)$ with the $N_s$, $N_c$, $N_{Alpha}$, $N_h$, $N_b$, and $N_g$, which are equal to i. These data were well fitted by linear function and the best fitting parameters values are listed in Table 6. From Figure 8 (a), the multiplicity of compound and shower particles had shown the strong correlation with normalized projectile's lightest fragments (proton). It can be seen that the normalized value of projectile's lightest fragments (proton) linearly increases with increasing number of shower particles ($N_s$). The normalized number of projectile's lightest fragments (proton) rapidly increases with the number of compound particle multiplicity ($N_c$) and after $N_c = 28$, it remarkably increases with the compound particles ($N_c$) multiplicity unlike the $<N_p>/N_p$ not remarkably increases with the number of alpha ($N_{Alpha}$) particles emitted in an event. From Figure 8 (b), the number of heavily ionizing charged particles ($N_h$), black particles ($N_b$) and the grey particles ($N_g$) have weak dependence with normalized multiplicity of projectile's lightest fragments (proton) from 1 to 9 but in this region normalized multiplicity of projectile's lightest fragments (proton) linearly increases with $N_h$, $N_b$, and $N_g$. It also shows that the number of $N_h$, $N_b$ and $N_g$ are equally interacting with H and CNO targets. Again the number of $N_h$ linearly increases with normalized multiplicity of projectile's lightest fragments (proton) beyond the value of 13 $N_h$ per event. It reflects that the heavily ionizing particles ($N_h$) not only equally interact with H and CNO target but are also strongly interacting with Ag/Br targets.

**5. Summary and Conclusion**.

We have significantly investigated the projectile's lightest fragments (proton) multiplicity distribution, probability distribution and its normalized multiplicity correlated with the secondary particles. From this investigation, we may conclude that, the mean multiplicity of the projectile's lightest fragments (proton) ($<N_p>$) completely depends on mass number of the projectile and does not showed significant depends on the projectile energy. The average multiplicity of the projectile's lightest fragments (proton) increases with increase in the target mass ($A_T$). The multiplicity distribution of projectile's lightest fragments (proton) emitted in the $^{84}Kr_{36}$- emulsion interaction well explained by the KNO scaling law and it is also obeying that law. The experimental data are well fitted with universal function and fitting values are almost close to the theoretical value within the statistical error. Correlation between $<N_h>$, $<N_b>$, $<N_g>$, $<N_s>$, $<N_{Alpha}>$, $<N_c>$ and projectile's lightest fragments (proton) $<N_p>$ had shown that, the average compound particles and shower particles showed strong correlation with projectile's lightest fragments (proton) $<N_p>$ rather than the black particles, grey particles, alpha particles and heavily ionizing charged particles. The correlation between $N_c$, $N_s$, $N_{Alpha}$, $N_b$, $N_g$, $N_h$ and $<N_p>$ shows that, the $<N_p>$ rapidly increases with increasing $N_c$, $N_s$ and $N_h$. In the case of $N_c$, the $<N_p>$ increases up to $N_C = 21$ after that $\langle N_p \rangle$ starts decreasing with increasing $N_c$ values. The $\langle N_p \rangle$ values are saturated for $N_h$ values beyond 13. The heavily ionizing charged particles and grey particles show similar strength on the projectile's lightest fragments (proton) within the statistical error. The slope values for $N_h$ and $N_g$ are (0.13±0.04) and (0.13±0.007), respectively. Normalized projectile's lightest fragments (proton) showed a strong correlation with the compound particles, shower particles and heavily ionizing charged particles rather than the black particles, grey particles and alpha particles. Normalized projectile's lightest fragment (proton) linearly increases with increasing $N_h$, $N_g$ and $N_b$ values from 1 to 9. It shows that $N_h$, $N_g$ and $N_b$ are equally interacting with targets H and CNO.

**Acknowledgment**

**Table1:** Gaussian distribution parameter values for different targets.

| Interactions | Peaks Value | Width |
|---|---|---|
| CNO ($N_h \geq 8$) | 2.013 | 6.816 |
| Kr - emulsion ($N_h \leq 0$) | -13.153 | 17.265 |

**Table 2:** Mean multiplicity of the projectile light fragment (proton) for different target groups and projectiles of similar and different kinetic energy.

| Projectile | Energy $A$ GeV | <$N_p$> | | | Reference |
|---|---|---|---|---|---|
| | | H | CNO | AgBr | |
| $^{12}$C | 3.7 | 0.77 ± 0.07 | 1.12 ± 0.04 | 0.86 ± 0.02 | [23] |
| $^{22}$Ne | 3.3 | 1.17 ± 0.02 | 1.47 ± 0.04 | 1.37 ± 0.03 | [21] |
| $^{56}$Fe | 1.0 | 2.53 ± 0.29 | 3.19 ± 0.28 | 4.94 ± 0.44 | [15] |
| $^{56}$Fe | 1.7 | 2.35 ± 0.17 | 3.00 ± 0.13 | 3.03 ± 0.09 | [22] |
| $^{84}$Kr | 1.7 | 2.44 ± 0.21 | 2.67 ± 0.15 | 3.06 ± 0.16 | [18] |
| $^{84}$Kr | 1.0 | 2.29 ± 059 | 3.55 ± 0.32 | 5.92 ± 0.62 | **Present work** |
| $^{132}$Xe | 1.0 | 1.9 ± 0.3 | 4.7 ± 0.3 | 6.7 ± 0.3 | [20] |

**Table 3:** The average multiplicities of projectile's light fragments (proton) for different projectile + emulsion inelastic interactions.

| Projectile | Energy (A GeV) | $<N_p>$ | Reference |
|---|---|---|---|
| $^4$He | 3.7 | 0.61 ± 0.02 | [16] |
| $^{12}$C | 3.7 | 1.20 ± 0.04 | [16] |
| $^{12}$C | 3.7 | 0.62 ± 0.03 | [19] |
| $^{12}$C | 3.7 | 0.93 ± 0.02 | [23] |
| $^{16}$O | 3.7 | 1.45 ± 0.03 | [16] |
| $^{16}$O | 3.7 | 1.15 ± 0.08 | [17] |
| $^{22}$Ne | 3.3 | 1.55 ± 0.02 | [16] |
| $^{22}$Ne | 3.3 | 1.36 ± 0.02 | [21] |
| $^{28}$Si | 3.7 | 2.54 ± 0.05 | [16] |
| $^{28}$Si | 3.7 | 1.82 ± 0.05 | [19] |
| $^{84}$Kr | 1.7 | 2.79 ± 0.10 | [18] |
| $^{84}$Kr | 1.0 | 3.55 ± 0.16 | **Present work** |
| $^{132}$Xe | 1.0 | 4.6 ± 0.2 | [20] |

**Table 4:** The values of the best fitting parameters of KNO scaling for $^{84}$Kr$_{36}$ - emulsion interactions along with the different targets at around 1 A GeV and 1.7 A GeV.

| Type of event | Energy (A GeV) | A | B | $\chi^2$ / DoF | Reference |
|---|---|---|---|---|---|
| $^{84}$Kr - H | 1.7 | 3.28 ± 0.959 | 2.10 ± 0.24 | 0.783 | [18] |
| $^{84}$Kr - CNO | 1.7 | 3.00 ± 0.507 | 1.95 ± 0.13 | 0.392 | [18] |
| $^{84}$Kr – Ag/Br | 1.7 | 3.46 ± 0.66 | 2.08 ± 0.16 | 1.140 | [18] |
| $^{84}$Kr - Em | 1.7 | 3.36 ± 0.35 | 2.02 ± 0.08 | 0.501 | [18] |
| $^{84}$Kr - H | 1.0 | 3.61 ± 0.6 | 1.93 ± 0.28 | 0.3091 | **Present work** |
| $^{84}$Kr - CNO | 1.0 | 3.7 ± 0.60 | 1.94 ± 0.28 | 0.321 | **Present work** |
| $^{84}$Kr – Ag/Br | 1.0 | 3.6 ± 0.60 | 1.95 ± 0.28 | 0.3176 | **Present work** |
| $^{84}$Kr - Em | 1.0 | 3.6 ± 0.59 | 1.93 ± 0.28 | 0.3091 | **Present work** |

**Table 5:** The best fitting parameter's values are listed for the $^{84}Kr_{36}$ - emulsion interaction at around 1 A GeV.

| Type of event | a | B |
|---|---|---|
| $<N_h>$ | 8.64 ± 0.97 | 0.15 ± 0.08 |
| $<N_b>$ | 4.72 ± 1.08 | 0.14 ± 0.09 |
| $<N_g>$ | 3.91 ± 0.57 | 0.04 ± 0.05 |
| $<N_s>$ | 5.19 ± 1.72 | 0.71 ± 0.15 |
| $<N_{Alpha}>$ | 1.23 ± 0.38 | 0.12 ± 0.03 |
| $<N_c>$ | 9.9 ± 2.33 | 1.08 ± 0.25 |

**Table 6:** The best fitting values of parameters for normalized multiplicity correlation for $^{84}Kr_{36}$ - Emulsion interactions at around 1 A GeV are listed.

| Type of event | a | B |
|---|---|---|
| $<N_s>$ | 0.0026 ± 0.0063 | 0.0046 ± 4.76×10$^{-4}$ |
| $<N_c>$ | 0.0287 ± 0.0068 | 0.0022 ± 3.622×10$^{-4}$ |
| $<N_{Alpha}>$ | -0.0405 ± 0.4425 | 0.0228 ± 0.0096 |
| $<N_h>$ | 0.0040 ± 0.0118<br>-0.0595 ± 0.0084 | 0.0066 ± 0.0017<br>0.0078 ± 4.6229×10$^{-4}$ |
| $<N_b>$ | 0.0077 ± 0.0041 | 0.0044 ± 5.1288×10$^{-4}$ |
| $<N_g>$ | -0.0109 ± 0.0170 | 0.0095 ± 0.0024 |

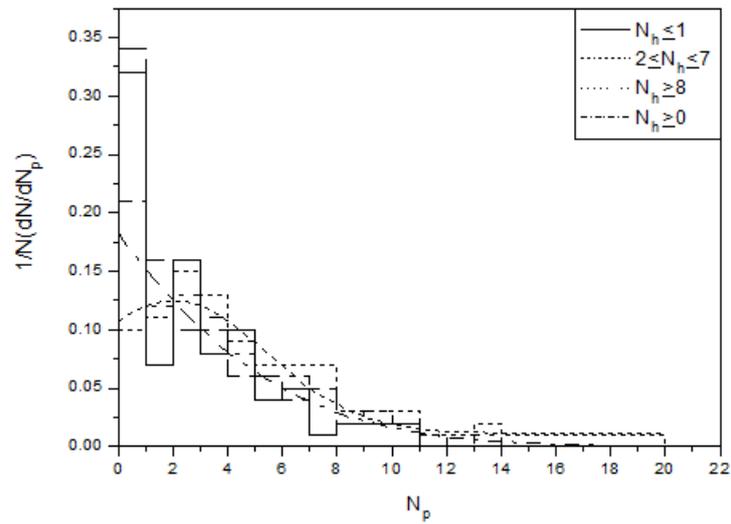

**Figure1:** The multiplicity distribution of projectile's lightest fragments (proton) ($N_p$) for the different target groups in the $^{84}Kr_{36}$ - emulsion interactions at around 1 A GeV.

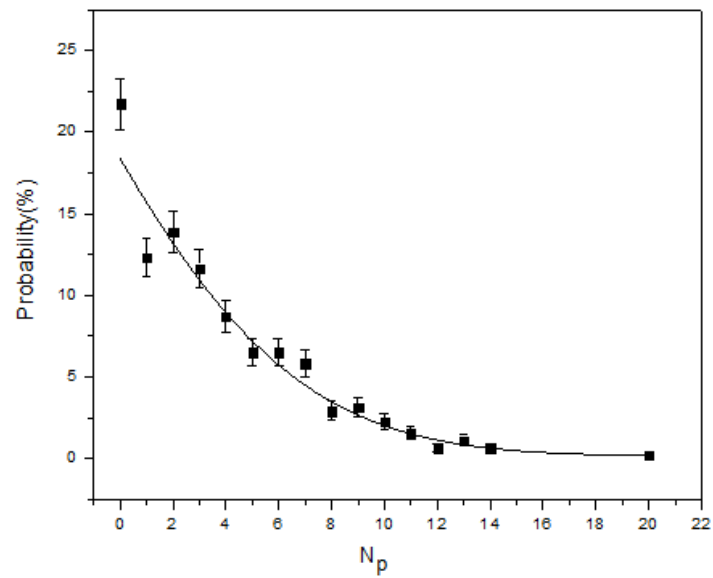

**Figure 2:** The probability distribution of the projectile's light fragment (proton) ($N_p$) for $^{84}Kr_{36}$ - emulsion interactions at around 1 A GeV.

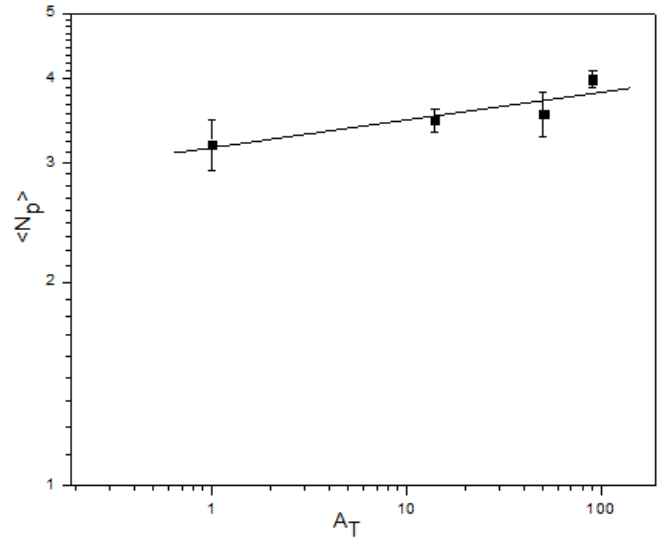

**Figure 3:** The average multiplicity of projectile's light fragments (proton) $N_p$ as a function of target mass number ($A_T$).

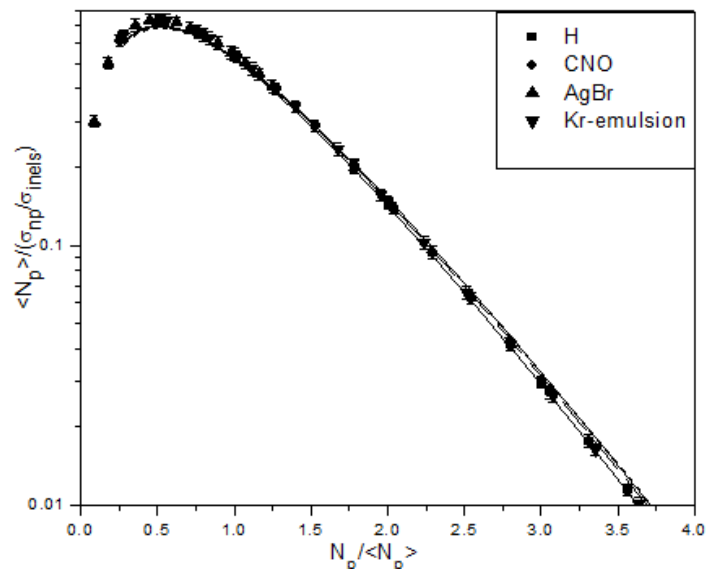

**Figure 4:** The KNO scaling distribution for the projectile's lightest fragments (proton) for the $^{84}Kr_{36}$ - emulsion interactions at around 1GeV per nucleon. The solid line represent the theoretical (KNO) fitting on the experimental data points ($N_p = 0$ excluded) with the function as shown in equation (4).

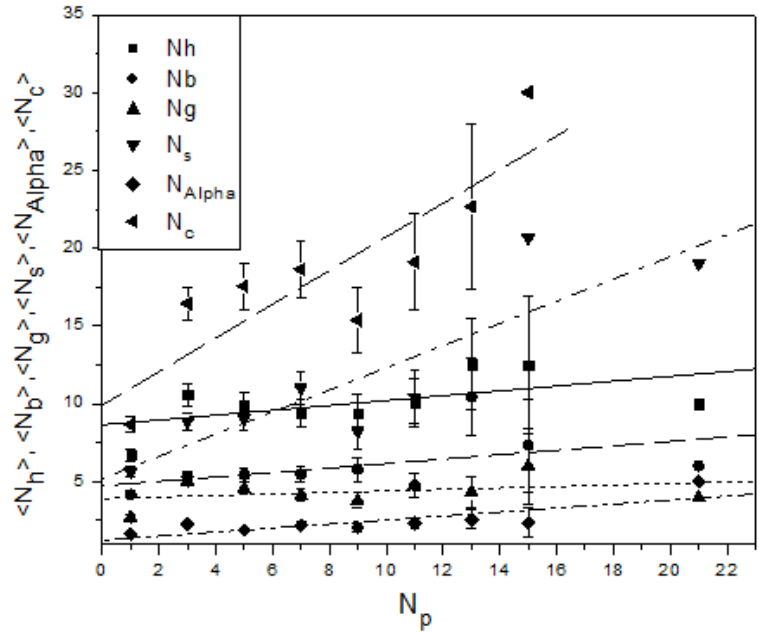

**Figure 5**: Multiplicity correlation between $\langle N_h \rangle$, $\langle N_b \rangle$, $\langle N_g \rangle$, $\langle N_s \rangle$, $\langle N_{Alpha} \rangle$, and $\langle N_c \rangle$ with $N_p$ for $^{84}Kr_{36}$ - emulsion interaction at around 1 A GeV.

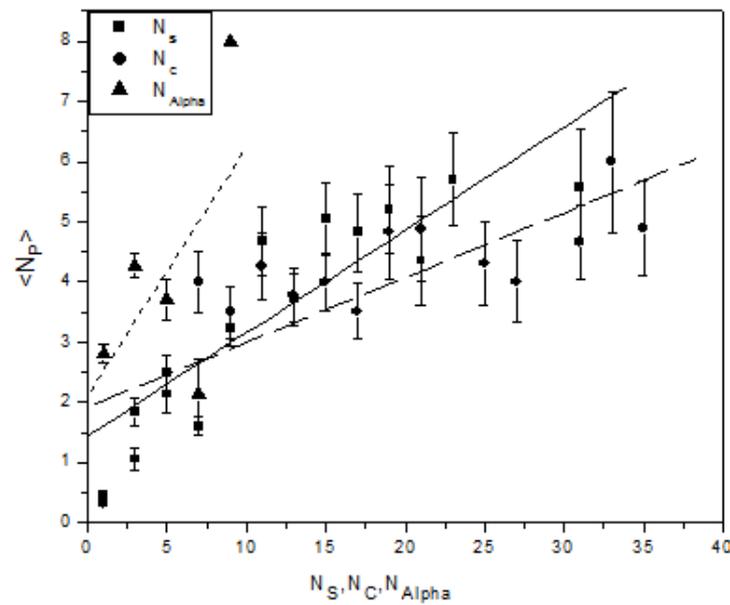

**Figure 6**: Multiplicity correlation between $\langle N_p \rangle$ with $N_s$, $N_c$, and $N_{Alpha}$ for $^{84}Kr_{36}$ - emulsion interactions at around 1 A GeV.

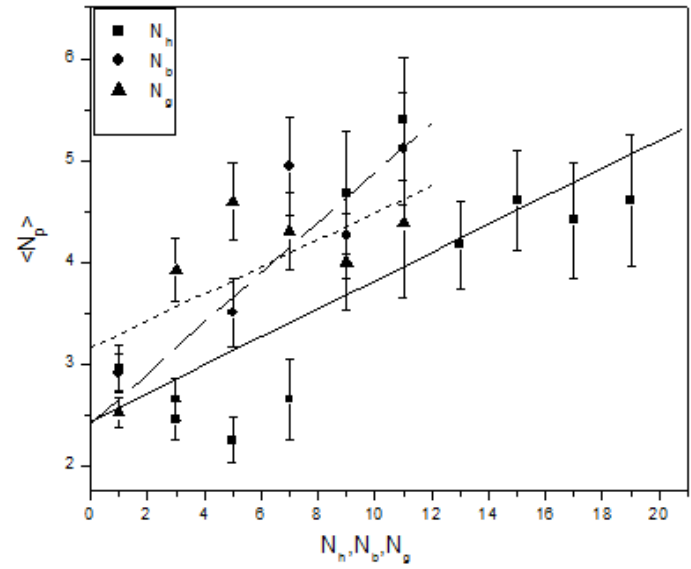

**Figure 7**: Multiplicity correlation between $<N_p>$ with $N_h$, $N_b$, and $N_g$ for $^{84}Kr_{36}$ - emulsion interactions at around 1 A GeV.

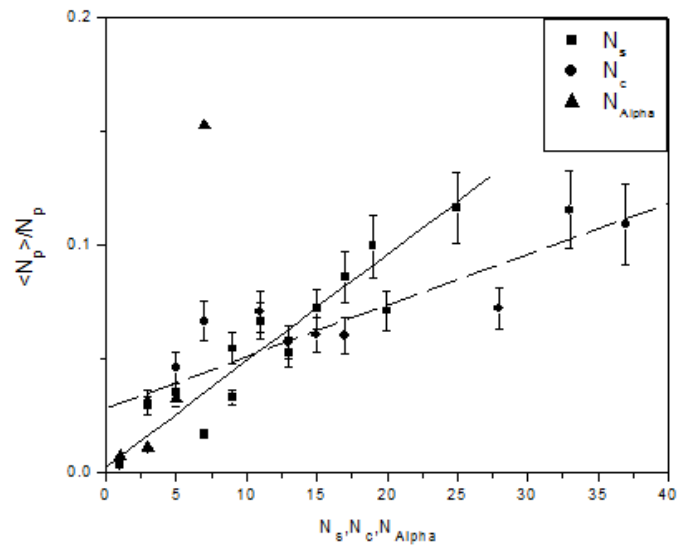

**Figure 8(a)**: Normalized multiplicity correlation between $<N_p>/N_p$ with the values of $N_s$, $N_c$, and $N_{Alpha}$ for $^{84}Kr_{36}$ - Emulsion interactions at around 1 A GeV.

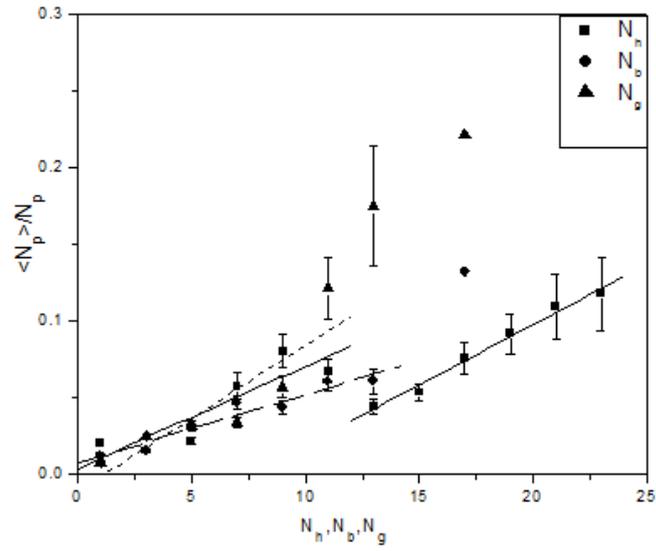

**Figure 8(b):** Normalized multiplicity correlation between $<N_p>/N_p$ with the values of $N_h$, $N_b$, and $N_g$ for $^{84}Kr_{36}$ - Emulsion interactions at around 1 A GeV.